\documentclass[
 ,final            
  ]
  {aipproc}

\layoutstyle{8s}

\usepackage{natbib}

\newcommand {\gsim}{\,\lower.7ex\hbox{$\;\stackrel{\textstyle>}{\sim}\;$}}
\newcommand {\lsim}{\,\lower.7ex\hbox{$\;\stackrel{\textstyle<}{\sim}\;$}}
\newcommand {\hi} {{\rm H}{\small\rm I}\,}

\begin{document}

\title[]{Galactic fountains and gas accretion}


\author{F. Marinacci}{
  address={Dipartimento di Astronomia, Universit\`a di Bologna, via Ranzani 1, 40127 Bologna, Italy.}
}

\author{J. Binney}{
  address={Rudolf Peierls Centre for Theoretical Physics, Oxford University, Keble Road, Oxford OX1
3NP, UK.}
}

\author{F. Fraternali}{
  address={Dipartimento di Astronomia, Universit\`a di Bologna, via Ranzani 1, 40127 Bologna, Italy.}
}

\author{C. Nipoti}{
  address={Dipartimento di Astronomia, Universit\`a di Bologna, via Ranzani 1, 40127 Bologna, Italy.}
}

\author{L. Ciotti}{
  address={Dipartimento di Astronomia, Universit\`a di Bologna, via Ranzani 1, 40127 Bologna, Italy.}
}

\author{P. Londrillo}{
  address={INAF-Osservatorio Astronomico di Bologna, via Ranzani 1, 40127,
Bologna, Italy.}
}

\begin{abstract}
Star-forming disc galaxies such as the Milky Way need to accrete $\gsim$ 1 $M_{\odot}$ of gas
each year to sustain their star formation. This gas accretion is likely to come from the cooling of the hot 
corona, however it is still not clear how this process can take place. We present simulations supporting the idea that 
this cooling and the subsequent accretion are caused by the passage of cold galactic-fountain clouds through the hot 
corona. The Kelvin-Helmholtz instability strips gas from these clouds and the stripped gas causes
coronal gas to condense in the cloud's wake. For likely parameters of the Galactic corona and of typical fountain clouds 
we obtain a global accretion rate of the order of that required to feed the star formation.
\end{abstract}

\maketitle


\section{The proposed scenario}
Star-forming disc galaxies like the Milky Way must accrete $\gsim$ 1 $M_\odot$ of
fresh gas each year \cite[see][and references therein]{Sancisi08} and have built their discs gradually 
over the last 10 Gyr \citep[e.g.][]{AumerB}.
A central question is the origin of the accreted gas and how this gas reaches the thin disc whitin which the process of 
star formation takes place.
The virial-temperature corona, in which disc galaxies are embedded, is the only reservoir of baryons capable of 
sustaining an accretion rate of $\sim 1\,M_{\odot}\,{\rm yr}^{-1}$ for a Hubble time.
We present the results of a set of grid-based hydrodynamical simulations supporting the idea that the gas needed by the 
disc to form stars is drawn from this corona.


Coronae of disc galaxies are similar in many respects to the hot
atmospheres of giant elliptical galaxies and galaxy clusters, but with
lower gas temperature and density
\cite[e.g.][]{Rasmussen09,Sembach03}. As the hot gas of these more massive systems, the coronal gas is unlikely to 
fragment into clouds via thermal instability \citep{Nipoti10}, but it is expected to cool monolithically and feed the 
central black hole rather than produce an extended cold disc in which stars can form.
However, if the gas needed to feed star 
formation has to be drawn from the corona, a 
mechanism that makes the hot gas accrete onto the disc must be at work.

There is abundant evidence that star formation in galaxies like the Milky Way powers a galactic fountain: ejection of gas from the mid-plane by supernova{} explosions  
\citep{HouckB90}.
Through the fountain a significant fraction (from 10 to 25 \%) of the whole \hi content of the galaxy is carried into its halo \cite[see][and references therein]{Fraternali10}.
In this work, supported by several lines of argument, we hypothesise that the transfer of gas from the corona to the star-forming disc is effected by the \hi clouds ejected by the galactic fountain \citep{Marinacci10}.

Our hydrodynamical
simulations suggest that the gas accretion proceeds through the following
steps: (i) stripping of gas from fountain clouds by the corona as a result of the Kelvin-Helmholtz instability, (ii) mixing of the (high metallicity) stripped gas with a comparable amount of coronal gas in the turbulent wake of the clouds; as a consequence of the mixing, the gas cooling time of the
coronal gas is reduced to a value lower than the cloud's flight time, (iii)  formation of knots of cold gas that accrete onto the disc in a dynamical
time.

\begin{figure}
\resizebox{0.8\columnwidth}{!}
  {\includegraphics{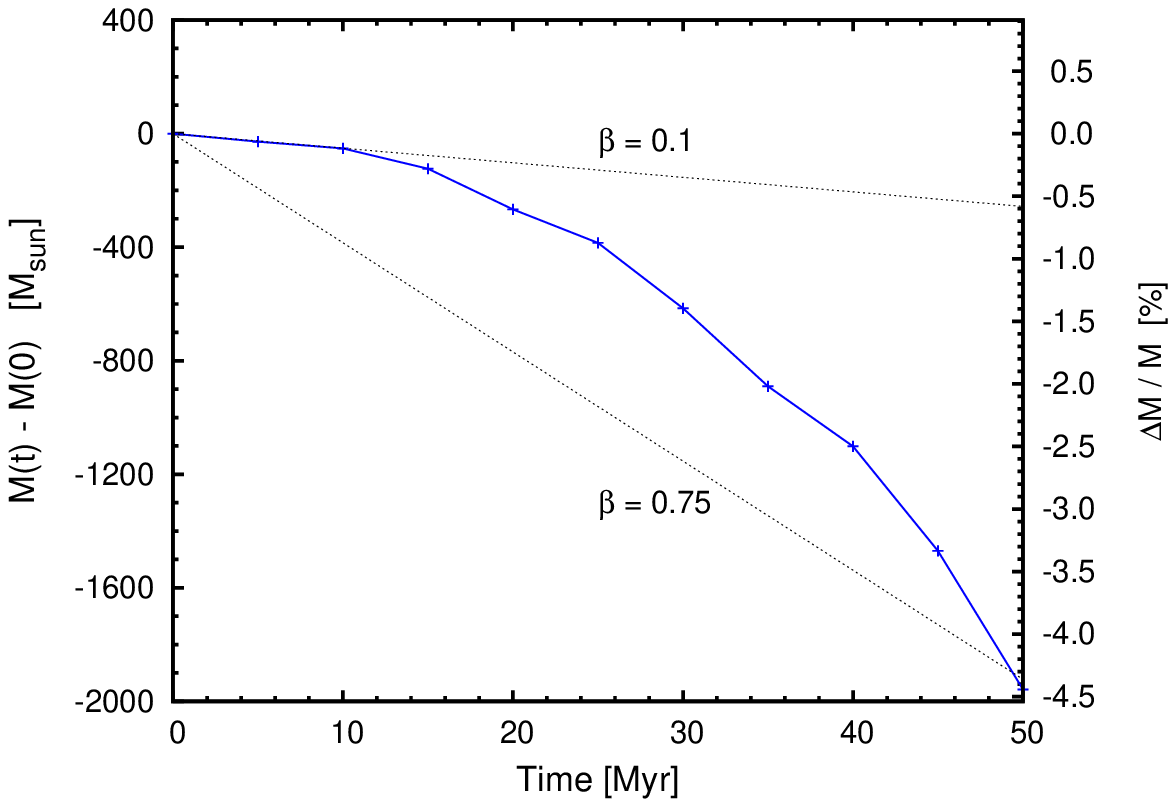}\includegraphics{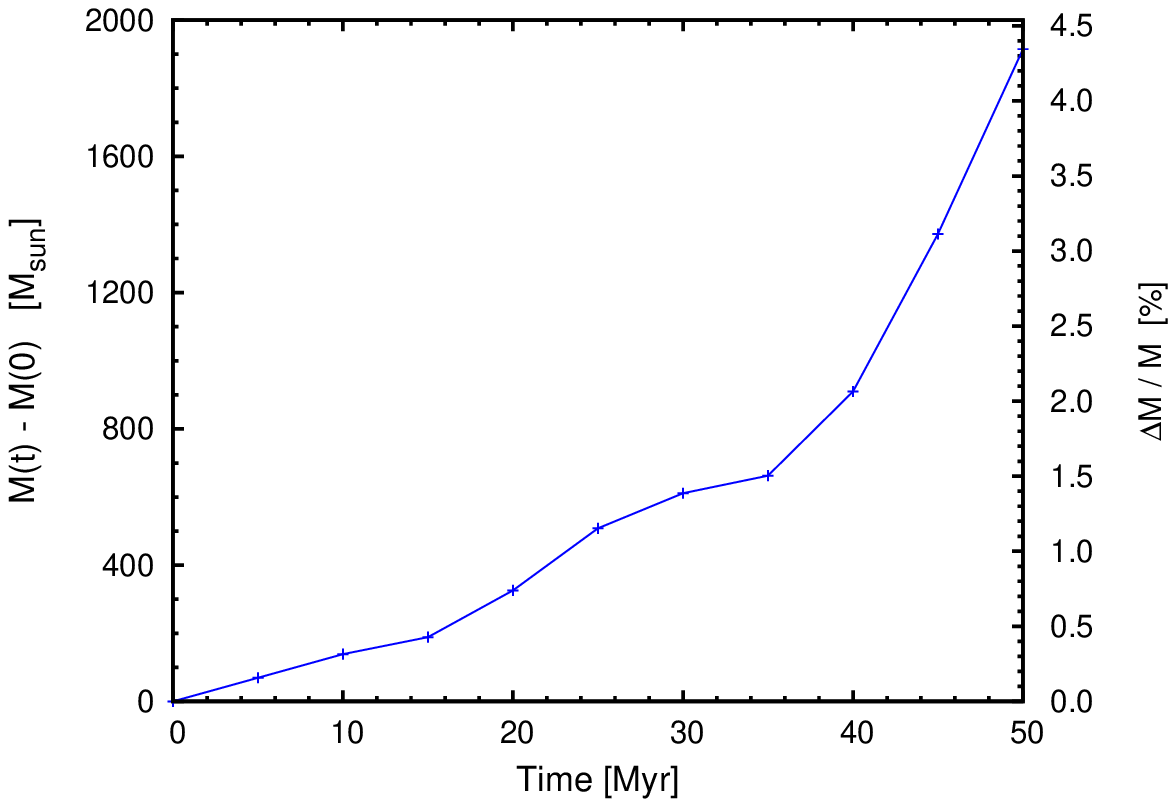}}
  \caption{The evolution of the mass of gas at $T<5\times10^5$ K when radiative cooling is switched off (left panel) or
  on (right panel). The dotted straight lines show two estimates of critical mass-loss rate $\alpha_{\rm crit}$.
 The particle density of the corona is $2 \times 10^{-3}{\rm cm}^{-3}$, its temperature $2 \times 10^{6}$ K and the 
 metallicity of the system is [Fe/H] $= -1.0$.}
  \label{nocool}
\end{figure}

The stripped gas leads to condensation of coronal gas only if the mass-loss rate exceeds a critical value $\alpha_{\rm crit}$, determined by the physical properties of the cloud and the corona. In addition,
dimensional analysis suggests that the actual mass-loss rate must lie close to
$\alpha_{\rm crit}$.
In view of the proposed scenario and of these
considerations the aims of the simulations are two-fold: (i) to provide an estimate of the actual mass-loss rate $\alpha$ for comparison with $\alpha_{\rm crit}$, (ii) to determine the critical ambient pressure (dependent on metallicity)
above which the mass of cool gas increases with time through
condensation in the wake.
In figure \ref{nocool} the evolution of the mass of gas at $T<5\times10^5$ K (accreted gas), for a simulation with physical conditions representative of galactic coron\ae{}, is shown (see caption for details). In particular, in the left panel of the figure, the mass loss of a cloud is displayed and the agreement between the simulation and the analytical prediction is remarkable. From the right panel of figure \ref{nocool} it has been possible to derive an estimate of the global accretion rate ($\approx 0.5 M_{\odot}\,{\rm yr}^{-1}$) which is of the same order as that required to feed star formation.
By comparing the figures in the two panels it is apparent that the cooled mass of
coronal gas is comparable to the mass lost by the cloud. In other words, when the cooling is switched on the evaporation 
of some cloud mass produces an accretion of roughly the same amount of coronal mass.

\section{Summary and conclusions}
We have used hydrodynamical simulations to check whether the interaction between the galactic-fountain clouds, powered by a star-forming disc, and the virial-temperature corona, in which disc galaxies are embedded, could lead the coronal gas to cool promptly in the clouds' wake and accrete onto the disc to feed star formation.
The results of our analysis, described in detail in \cite{Marinacci10}, can be summarized as follows:
\begin{itemize}
\item the interaction between the galactic fountain and the hot corona causes the fountain clouds to be stripped of some of their gas. The simulations provide a reliable estimate of the mass-loss rate and confirm that its value lies close to $\alpha_{\rm crit}$, in agreement with analytical expectations;
\item  the condensation of coronal gas in the cloud's wake prevails over evaporation if the metallicity and/or pressure of the corona are high enough. In our simulations this happens for likely parameters of the coron\ae{} in disc galaxies,
condensation is present also for densities as low as $n_{\rm h} \simeq 4\times10^{-4} {\rm cm}^{-3}$, provided that 
[Fe/H] $\sim 0$;
\item  the derived global accretion rate is of the same order as that required to feed star formation. Therefore the  
condensation of the hot corona seems a viable
mechanism to sustain the star formation rate in star-forming galaxies.
\end{itemize}



\begin{theacknowledgments}
We used CPU time assigned under the INAF-CINECA agreement 2008-2010. F. Marinacci gratefully acknowledges the support from the Marco Polo program by University of Bologna and the hospitality of the Rudolf Peierls Centre for Theoretical Physics.
\end{theacknowledgments}

\end{document}